\documentclass[12pt,preprint]{aastex}
\usepackage{color}
\shorttitle{VIRGO-SHEAR ALIGNEMENTS}
\shortauthors{Lee et al.}
\begin{document}
\title{ALIGNMENTS OF THE GALAXIES IN AND AROUND THE VIRGO CLUSTER WITH THE LOCAL VELOCITY SHEAR }
\author{Jounghun Lee\altaffilmark{1}, 
Soo Chang Rey\altaffilmark{2}, and Suk Kim\altaffilmark{2}}
\altaffiltext{1}{Astronomy Program, Department of Physics and Astronomy, Seoul National University, 
Seoul 151-747, Korea\\ 
\email{ jounghun@astro.snu.ac.kr}}
\altaffiltext{2}{Department of Astronomy and Space Science, Chungnam National University,
Daejeon 305-764, Korea}
\begin{abstract} 
An observational evidence is presented for the alignment between the cosmic sheet and the principal axis of the velocity shear 
field at the position of the Virgo cluster. The galaxies in and around the Virgo cluster from the Extended Virgo Cluster Catalog 
recently constructed by Kim et al. are used to determine the direction of the local sheet. 
The peculiar velocity field reconstructed from the Sloan Digital Sky Survey Data Release 7 is analyzed to estimate the local 
velocity shear tensor at the Virgo center. Showing first that the minor principal axis of the local velocity shear tensor is almost parallel to the 
line of sight direction,  we detect a clear signal of alignment  between the positions of the Virgo satellites and the 
intermediate principal axis of the local velocity shear projected onto the plane of the sky.  Furthermore, the dwarf satellites 
are found to appear more strongly aligned than the normal counterparts, which is interpreted as indication of the following: 
(i) The normal and the dwarf satellites fall in the Virgo cluster preferentially along the local filament and the local sheet, respectively. 
(ii) The local filament is aligned with the minor principal axis of the local velocity shear while the local sheet is in parallel to the plane 
spanned by the minor and the intermediate principal axes.
Our result is consistent with the recent numerical claim that the velocity shear is a good tracer of the cosmic web.
\end{abstract}
\section{INTRODUCTION}

According to the density peak formalism \citep{bbks86},  the formation of a galaxy cluster occurs at a rarely high peak of the linear 
density field where the three eigenvalues of the density Hessian have almost identical values. Since the ellipticity and the prolateness 
of an initial density region are expressed in terms of the differences among the three eigenvalues of its density Hessian, 
the geometrical shape of a proto-cluster  located at a high density peak is expected to be quite spherically symmetric. 
Moreover, \citet{bernardeau94} have analytically showed  with the help of the linear perturbation theory that the collapse process 
of a rare density peak should follow well the spherical dynamics at first order. Several numerical experiments supported this 
analytic prediction:  For instance, the values of the linear density contrast $\delta$ in the proto-cluster regions were found by 
\citet{robertson-etal09} to converge to the spherical threshold value $\delta_{c}=1.69$ predicted by the spherical dynamical model 
\citep[e.g.,][]{peebles80}. The mass function of cluster-size dark halos determined from $N$-body simulations showed good 
agreements with the analytic prediction based on the spherical excursion set theory \citep{MR10}. 

Meanwhile, the optical, the X-ray and the Sunyaev-Zel'dovich  (SZ) effect observations of the galaxy clusters have demonstrated 
for long that the shapes of the galaxy clusters must be far from being spherically symmetric 
\citep[e.g.,][]{binggeli82,fabricant-etal84,BC96,kawahara10,sayers-etal11}, 
which was also confirmed by the weak lensing analyses of the matter density profiles of the galaxy clusters \citep[e.g.,][]{EB09,oguri-etal10}. 
The non spherical shapes of the rarest bound objects have also been noted by $N$-body experiments. For instance,
very recently, \citet{despali-etal14} employed an ellipsoidal halo finder algorithm to a high-resolution $N$-body simulation and 
revealed that  the more massive a dark halo is, the less spherical its shape is. 
For a comprehensive review on the non-spherical shapes of galaxy clusters, see \citet{clustershape_review13}. 

To explain how the galaxy clusters develop non-spherical shapes even though they originate from the spherically symmetric proto-cluster sites, 
the "anisotropic merging scenario" has been proposed and prevalently accepted \citep[e.g.,][]{west-etal91,VV93,west-etal95}, according to which 
the preferential occurrence of the merging and infall of satellite galaxies along the "cosmic web" \citep{web96} are responsible for the 
nonspherical shapes of the galaxy clusters. This anisotropic merging scenario makes three key predictions. 
First, the cluster satellites tend to be aligned with the directions of the cosmic web that interconnects the clusters to the 
surrounding large-scale structure \citep{ara-etal07a,hahn-etal07b,zhang-etal09,codis-etal12,for-etal14}. 
Second, the satellite galaxies around the dynamically younger clusters are more strongly aligned with the cosmic web than the relaxed ones.  
Third, those clusters located in the filament or sheet environments have more elongated shapes than those in the knot or void environments 
since the merging of their satellite galaxies occurs more anisotropically in the filament and sheet environments. 

The alignments between the shapes of dark matter halos and the directions of the cosmic web induced by the anisotoropic merging have 
been investigated by many numerical studies \citep[e.g.,][]{ara-etal07a,hahn-etal07b,zhang-etal09,codis-etal12,for-etal14}. However, 
these investigations had one difficulty that there is no unique web-classification scheme.
A variety of algorithms have been developed to classify the cosmic web, each of which has its own advantage and disadvantage: 
a method based on the Minimal Spanning Tree algorithm \citep{colberg07}, the Skeleton algorithm \citep{nov-etal06,sou-etal08}, the 
Multiscale Morphology Filter algorithm \citep{ara-etal07b}, the schemes based on the tidal shear 
\citep[e.g.,][]{hahn-etal07a,for-etal09}, density Hessian \citep[e.g.,][]{bond-etal10}, and velocity shear 
\citep{hoffman-etal12,libeskind-etal13a},  the SpineWeb method \citep{ara-etal10}, the NEXUS algorithm \citep{nexus13} and so on. 
For a concise review of those various web-classification algorithms, see \citet{nexus13}.

Here, we employ the web-classification scheme based on the velocity shear field proposed by \citet{hoffman-etal12}. The knots, filaments, 
and sheets classified from $N$-body simulations according to their scheme were shown to resemble very well the zero, one and two 
dimensional structures, respectively. In the subsequent works, \citet{libeskind-etal13a} applied the web-classification of \citet{hoffman-etal12} 
to their $N$-body data  and found that there exist strong alignments between the most elongated axes of the triaxial dark matter halos and the 
minor principal axes of the velocity shear tensors. The $N$-body results obtained by \citet{tempel-etal14} also revealed that  the minor 
principal axes of the velocity shear field trace well the directions of the large-scale filaments. 

Here, our goal is to observationally test the numerical claim that the directions of the cosmic web along which the satellite galaxies fall into the 
clusters are aligned with the principal axes of the local velocity shear. 
This observational test will be optimized by using a dynamically young cluster with a conspicuously elongated shape located in 
a filament or a sheet environment. In this respect, the nearby Virgo cluster is an appealing test-bed since it has yet to reach a dynamical 
equilibrium \citep[e.g.,][]{binggeli-eta87}, having a conspicuously non-spherical shape elongated along the direction almost parallel to the line 
of the sight  \citep[e.g.,][and references therein]{WB00,gavazzi-etal99,mei-etal07}, located in the Local Supercluster 
\citep[see][and references therein]{tully82}. The previous works which studied the three dimensional shapes of the Virgo cluster and its alignments 
with the local filaments focused largely on the central bright galaxies within the Virgo's virial radius. Very recently, \citet{kim-etal14} constructed an 
unprecedentedly extensive catalog which contains even faint dwarf galaxies located in the regions well outside the virial radius, which must 
be very useful for the study of the connection between the spatial distribution of the Virgo satellites and the local velocity shear field.

The outlines of the upcoming sections are as follows. In section \ref{sec:vshear} we explain how the velocity shear tensor at the Virgo center is 
determined from the peculiar velocity field reconstructed from the Sloan Digital Sky Survey Data Release 7 (SDSS DR7) \citep{sdssdr7}. 
In section \ref{sec:align} we measure the spatial distributions of the galaxies in and around the Virgo cluster and investigate their alignments 
with the principal axes of the local velocity tensors. In section \ref{sec:con} we summarize the result and discuss its physical implication. 

\section{RECONSTRUCTION OF THE VELOCITY SHEAR FIELDS}\label{sec:vshear}

\citet{hoffman-etal12} defined the normalized velocity shear field ${\bf \Sigma}\equiv (\Sigma_{ij})$ by 
\begin{equation}
\label{eqn:vshear}
\Sigma_{ij}=-\frac{1}{2H_{0}}\left(\partial_{j}{v_{i}} + \partial_{i}{v_{j}}\right)\, ,
\end{equation}
where ${\bf v}\equiv (v_{i})$ is the comoving peculiar velocity vector  smoothed on a scale of $R_{f}$ and 
$H_{0}$ is the Hubble constant. 
Letting the three eigenvalues of ${\bf \Sigma}$ be $\lambda_{1},\ \lambda_{2},\ \lambda_{3}$ with 
$\lambda_{1}\ge\lambda_{2}>\lambda_{3}$, they classified the regions with $\lambda_{3}>\lambda_{\rm th}$ as the knot 
environments, the regions with $\lambda_{2}>\lambda_{\rm th}>\lambda_{3}$ as the filament environments,  the regions with 
$\lambda_{1}>\lambda_{\rm th}>\lambda_{2}$ as the sheet environments and the regions with $\lambda_{1}<\lambda_{\rm th}$ as 
the void environments, where $\lambda_{\rm th}$ is an arbitrary threshold.  \citet{hoffman-etal12} empirically determined this 
threshold value to be $\lambda_{\rm th}=0.44$ which yielded the best visualisation of the geometrical characteristics of the four 
environments at  $z=0$ for a flat $\Lambda$CDM ($\Lambda$ dominated with cold dark matter) cosmology. 
Note that this threshold value $\lambda_{\rm th}$ is likely to depend on the background cosmology as well as on the redshift.
 
As mentioned in \citet{hoffman-etal12}, on the linear scale where the peculiar velocity field is curl free, the velocity shear field 
in Equation (\ref{eqn:vshear}) is identical to the tidal shear field defined as the second derivative of the gravitational potential field. 
Whereas, on the nonlinear scale where the peculiar velocity field is no longer curl free, the velocity shear field deviates from 
the tidal shear field. Very recently, however, \citet{libeskind-etal14} found from a $N$-body experiment that the directions of the 
principal axes of the velocity shear field are insensitive to the smooting scales, showing only slight changes when $R_{f}$ varies from 
the nonlinear scale $0.125\,h^{-1}$Mpc to the linear scale of $8\,h^{-1}$Mpc. Their numerical result implied that the directions of the 
principal axes of the real velocity shear field could be well approximated by those of the velocity shear field linearly reconstructed 
under the assumption that the peculiar velocity field is curl free.

To determine the local velocity shear tensor at the position of the Virgo cluster by Equation (\ref{eqn:vshear}), we use the peculiar 
velocity field that \citet{wang-etal12} have linearly reconstructed on a cubic volume of $726^{3}\,h^{-3}$Mpc$^{3}$ by applying the 
{\it halo-tracing} algorithm developed by \citet{wang-etal09} to the New-York University Value-Added Galaxy Catalog 
\citep{blanton-etal05} from SDSS DR7 \citep{sdssdr7}.
Dividing the volume into a total of $494\times 892 \times 499$ cubic grids each of which has a linear size of $0.71\,h^{-1}$Mpc, 
\citet{wang-etal12} statistically estimated the expectation value of the peculiar velocity vector on each grid, assuming 
a WMAP5 cosmology \citep{wmap5}. For a detailed description of how the peculiar velocity field was reconstructed from the SDSS 
DR7, see \citet{wang-etal12}.

Since the peculiar velocities on those grids near the limits of the SDSS DR7 volume are likely to have 
been contaminated by the boundary effect \citep{wang-etal09,wang-etal12}, we consider only the innermost subvolume of 
$\sim 180^{3}\,h^{-3}$Mpc$^{3}$ consisting of $256^{3}$ grids located well inside the boundary of the SDSS DR7 volume 
for the reconstruction of the velocity shear field.  The Cartesian coordinates of the selected grids in the equatorial reference frame 
are in the range of $-182.28\le x_{\rm grid}/(h^{-1}{\rm Mpc})\le\,-1.49,\ -90.05\le y_{\rm grid}/(h^{-1}{\rm Mpc})\le\,-90.74,\ 
-22.73\le z_{\rm grid}/(h^{-1}{\rm Mpc})\le\,158.07$.

Performing the Fourier transformation of the peculiar velocity field by employing the Fast Fourier Transformation (FFT) method 
\citep{press-etal92}, we obtain its Fourier amplitude field, $\tilde{\bf v}$.  Since the peculiar velocity field reconstructed by 
\citet{wang-etal12} is an unsmoothed version, we apply a Gaussian window function with a filtering radius of $R_{f}=1\,h^{-1}$Mpc 
to $\tilde{\bf v}$. The Fourier amplitude of the velocity shear field smoothed on a scale of $R_{f}$, 
$\tilde{\bf \Sigma}\equiv (\tilde{\Sigma}_{ij})$, can be obtained as 
\begin{equation}
\label{eqn:vshear_f}
\tilde{\Sigma}_{ij} = -\frac{1}{2H_{0}}\left(k_{j}\tilde{v}_{i} + k_{i}\tilde{v}_{j}\right)\exp\left(-\frac{R_{f}\tilde{v}^{2}}{2}\right) \, .
\end{equation}
Performing the inverse Fourier transform of $\tilde{\bf \Sigma}$ with the help of the inverse FFT method 
\citep{press-etal92}, we pull it off reconstructing the smoothed velocity shear field, ${\bf \Sigma}({\bf x})$ on the subvolume.

\begin{figure}[ht]
\begin{center}
\epsscale{1.0}
\plotone{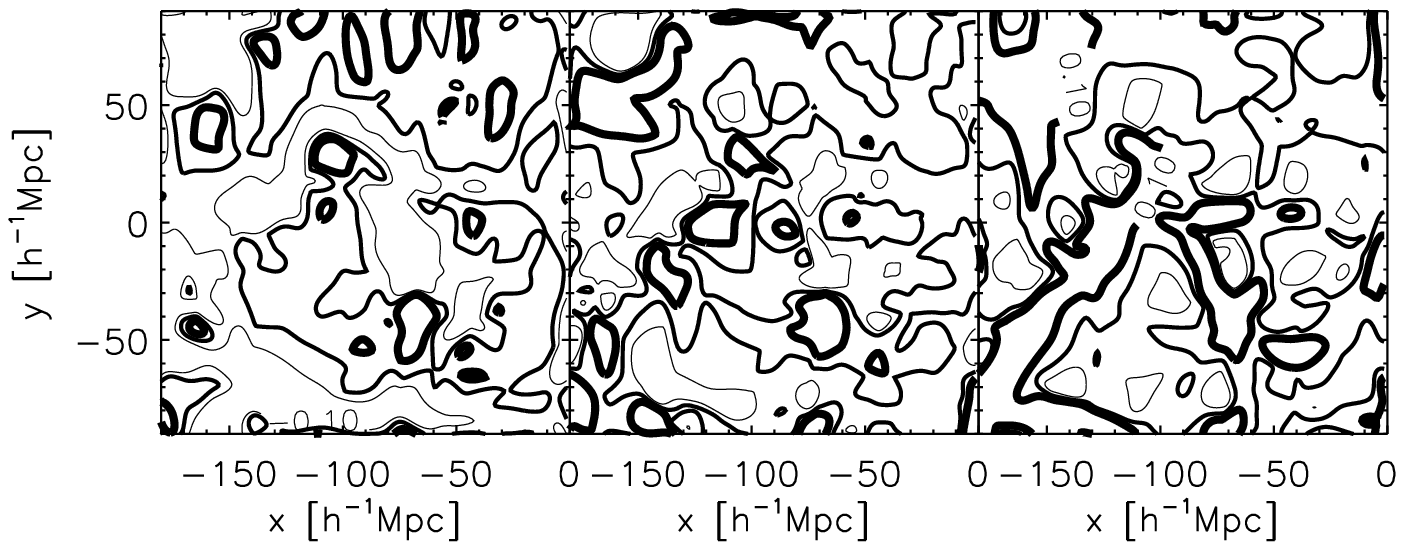}
\caption{Projected contours of the smallest ($\lambda_{3}$), second to the largest ($\lambda_{2}$), 
and largest ($\lambda_{1}$) eigenvalue of the velocity shear field smoothed on the scale of $R_{f}=5\,h^{-1}$Mpc 
for three different contour levels in the left, middle and right panel, respectively:  $\lambda_{3}=(-0.1,\ -0.05,\ 0.0)$; 
$\lambda_{2}=(-0.05,\ 0.0,\ 0.05)$; $\lambda_{1}=(0.0,\ 0.05,\ 0.1)$. In each panel,  
the contour level in a decreasing order is represented by the thickest, thick and thin line, respectively.}
\label{fig:contour}
\end{center}
\end{figure}

The equatorial  Cartesian coordinates of the Virgo center that almost coincides with the galaxy M87 have been found to be 
$x=-12.66\,h^{-1}$Mpc, $y=-1.71\,h^{-1}$Mpc, $z=2.81\,h^{-1}$Mpc \cite[e.g.,][]{mei-etal07}. 
Locating the grid nearest to the position of the Virgo center in the subvolume and calling it the Virgo grid, we find the local velocity 
shear tensor at the Virgo grid and diagonalize it via the similarity transformation to find its three eigenvectors and the corresponding 
eigenvalues. The major, the intermediate, and the minor principal axis of the local velocity tensor estimated on the Virgo grid is 
determined as its eigenvectors corresponding to the largest, the second to the largest and the smallest eigenvalue, respectively. 
It is found that two eigenvalues are positive while the smallest one is negative at the Virgo grid, which indicates the Virgo cluster is 
located in the filamentary environment according to the web-classfication scheme of \citet{hahn-etal07a}. Figure \ref{fig:contour} plots 
the contours of the smallest (left panel), second to the largest (middle panel), and largest eigenvalue (right panel) of the velocity shear 
field smoothed on the scale of $R_{f}=5\,h^{-1}$Mpc in the projected $x$-$y$ plane. 

We measure the orientation angles of the three principal axes of the local velocity shear tensor at the Virgo grid relative to the line 
of sight direction to the Virgo center and find the major, the intermediate, and the minor principal axis to be inclined $94.7^{\rm o}$, 
$98.7^{\rm o}$ and $9.9^{\rm o}$ from the direction of the sight line to the Virgo center, respectively. This result explicitly shows that 
the minor principal axis of the local velocity shear at the Virgo grid is well aligned with the line of sight direction while the other two 
principal axes tend to lie in the plane of the sky perpendicular to the sight line. Given the observational evidence found by \citet{WB00} 
that the elongated axis of the spatial distribution of the central bright galaxies in the Virgo cluster points almost toward the line of the 
sight direction \citep[see also][]{gavazzi-etal99,mei-etal07}, this result indicates that the longest axis of the Virgo shape determined 
by the spatial distribution of the central bright galaxies is well aligned with the minor principal axis of the local velocity shear. 

Provided that the spatial distribution of the central bright galaxies in the Virgo cluster is indeed elongated with the local filament as 
claimed by \citet{WB00},  this result observationally supports the numerical predictions of \citet{libeskind-etal13a} and 
\citet{tempel-etal14} that the cosmic filaments are in the direction of the minor principal axes of the velocity shear fields. 
Yet, it is not guaranteed that the elongated axis of the spatial distribution of the central bright galaxies located within the virial 
radius of the Virgo cluster represents the direction of the local filament.  It is definitely necessary to consider more galaxies in and 
around the Virgo cluster for the determination of the local direction of the cosmic web that connects the Virgo cluster to the surrounding 
large scale structure and for the measurement of its correlation with the principal axes of the local velocity shear. 

\section{ANISOTROPIC DISTRIBUTION OF THE  VIRGO SATELLITES}\label{sec:align}

To find an observational evidence for the alignments between the spatial distributions of the galaxies in and around the Virgo cluster 
and the principal axes of the local velocity shear tensor determined in section \ref{sec:vshear}, we utilize the Extended Virgo Cluster 
Catalog (EVCC) constructed by \citet{kim-etal14} which contains a total of $1589$ galaxies with the $r$-magnitudes less than $17.77$ 
mag corresponding to $M_{r}$ = $-13.33$ mag at the Virgo cluster distance of 16.5 Mpc (i.e., a distance modulus 
$m-M$ = 31.1 mag) \citep{jerjen-etal04,mei-etal07}. 
These galaxies are located in the regions of $750$ square degree centered around the Virgo cluster.
Information on the spectroscopic properties of the EVCC galaxies was extracted mainly from the SDSS DR7 and also from the NASA 
Extragalactic Database (NED). For a  detailed description of the EVCC, see \citet{kim-etal14}.

\begin{figure}[ht]
\begin{center}
\epsscale{0.6}
\plotone{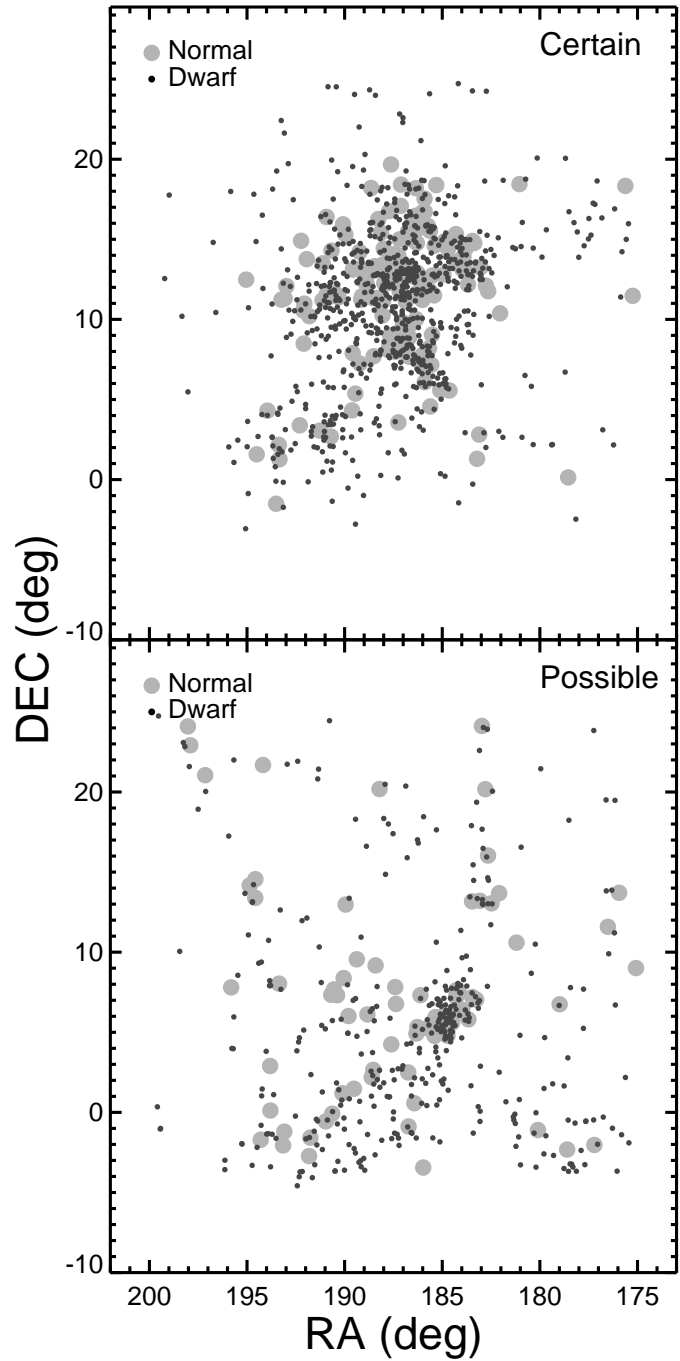}
\caption{A map of the equatorial positions of the certain and possible members from the EVCC \citep{kim-etal14} in the 
top and the bottom panel, respectively. In each panel, the black small (grey large) dots correspond to the dwarf 
(normal) satellites. }
\label{fig:2dmap}
\end{center}
\end{figure}

The EVCC galaxies were divided by \citet{kim-etal14} into the certain and the possible members of the Virgo cluster according to
their heliocentric radial velocities, $v_{l}\hat{\bf l}_{c}$, as a function of the separation distance $r_{2d}$ from the Virgo center in the 
plane of the sky where $\hat{\bf l}_{c}$ denotes the unit vector in the radial direction. The certain members are the dynamically 
relaxed central galaxies of the Virgo cluster that satisfy the criterion of 
$-v_{c}({\bf r}_{2d})\le v_{l}({\bf r}_{2d})\le v_{c}({\bf r}_{2d})$ where $v_{c}({\bf r}_{2d})$ denotes the threshold of the magnitude of the 
infall velocity at ${\bf r}_{2d}$ prescribed by the cluster infall model \citep{PS94}. 
Whereas, the possible members in the EVCC which do not meet the above criterion could be the satellite galaxies around the 
Virgo cluster, being located in the cosmic web that encompasses the Virgo and the surrounding large scale structures. The EVCC 
includes a total of $1028$ ($561$) certain (possible) members \citep{kim-etal14}.
The top and bottom panel of Figure \ref{fig:2dmap} shows the spatial distributions of the certain and possible members, respectively, in the 
equatorial coordinate frame (right ascension and declination). In each panel, the black small (grey large) dots correspond to 
the dwarf (normal) EVCC galaxies. 

The three dimensional positions of the EVCC galaxies relative to the Virgo center are hard to measure accurately due to the presence 
of the nonlinear redshift distortion effect \citep[see][for a review]{hamilton98}.  
To avoid the uncertainty caused by the inaccurate measurements of the positions of the Virgo satellites, we project the position vectors 
of the EVCC galaxies and the principal axes of the local velocity tensor reconstructed in section \ref{sec:vshear} onto the plane of the 
sky and measure the alignments between the projected directions. Although this process of the projection is expected to diminish the 
strength of an alignment signal (if any), this is the most secure way to avoid the effect of the nonlinear redshift distortion.

Using the equatorial coordinates, we first measure the position vector of each EVCC galaxy, ${\bf r}$, relative to the Virgo center, 
M87, under the assumption that it is at the same distances from us as the Virgo center. For the determination of the galaxy positions, 
we assume a WMAP5 cosmology \citep{wmap5} with $\Omega_{m}=0.258$, $\Omega_{\Lambda}=0.742$, $h=0.72$, to be consistent with 
\citet{wang-etal12}. The two dimensional position vector, ${\bf r}_{2d}$, of each EVCC galaxy is then obtained  as 
${\bf r}_{2d}={\bf r}-({\bf r}\cdot\hat{\bf l}_{c})\hat{\bf l}_{c}$. 
In a similar manner, we also project the three principal axes of the local velocity tensor at the Virgo grid onto the plane of the sky. 
Let ${\bf p}_{2d,1},\ {\bf p}_{2d,2}$ and ${\bf p}_{2d,3}$ denote the projected major, intermediate, and minor principal axes of the local 
velocity shear tensor, respectively. In section \ref{sec:vshear}, it has been already found that the minor principal axis of the local velocity 
shear is well aligned with ${\bf l}_{c}$. Thus, the projection would not affect much the directions of the major and the intermediate axes of the 
local velocity shear while the true direction of the minor principal axis must be completely lost by the projection.

We measure the angle, $\theta_{2d}$, between the projected principal axes of the local velocity shear tensor and the 
projected position vector of each EVCC galaxy as
\begin{equation}
\label{eqn:theta_2d} 
\theta_{2d}\equiv \cos^{-1}\left(\frac{\vert{\bf r}_{2d}\cdot{\bf p}_{2d,i}\vert}{\vert{\bf r}_{2d}\vert\vert{\bf p}_{2d,i}\vert}\right),\  
\quad  i=1,\ 2,\ 3,\ 
\end{equation} 
where the range of $\theta_{2d}$ is restricted to $[0, \ \pi/2]$ since what matters is only the relative angles between the two directions 
rather than their actual orientations. 
Binning the values of $\theta_{2d}$ and counting the numbers of the EVCC galaxies, we determine the probability 
density distribution of the alignment angle as $p(\theta_{2d})\equiv \Delta N_{\rm sat}/(\Delta\theta_{2d}\cdot N_{\rm tot})$ where 
$\Delta\theta_{2d}$ represents the bin size, $N_{\rm tot}$ is the total number of the EVCC galaxies, and 
$\Delta N_{\rm sat}$ represents the numbers of the EVCC galaxies whose alignment angles lie in the differential range of 
$[\theta_{2d},\ \theta_{2d}+\Delta\theta_{2d}]$. We also determine $p(\theta_{2d})$ separately for the certain members and for the 
possible members. If the projected positions of the EVCC galaxies are preferentially aligned with  (perpendicular to) one of the principal 
axes of the local velocity shear, then the probability density distribution, $p(\theta_{2d})$, will decrease  (increase) as $\theta_{2d}$ 
increases. If no alignment, then $p(\theta_{2d})$ will have a constant value of $2/\pi$ over the whole range of $\theta_{2d}$.

\begin{figure}[ht]
\begin{center}
\epsscale{1.0}
\plotone{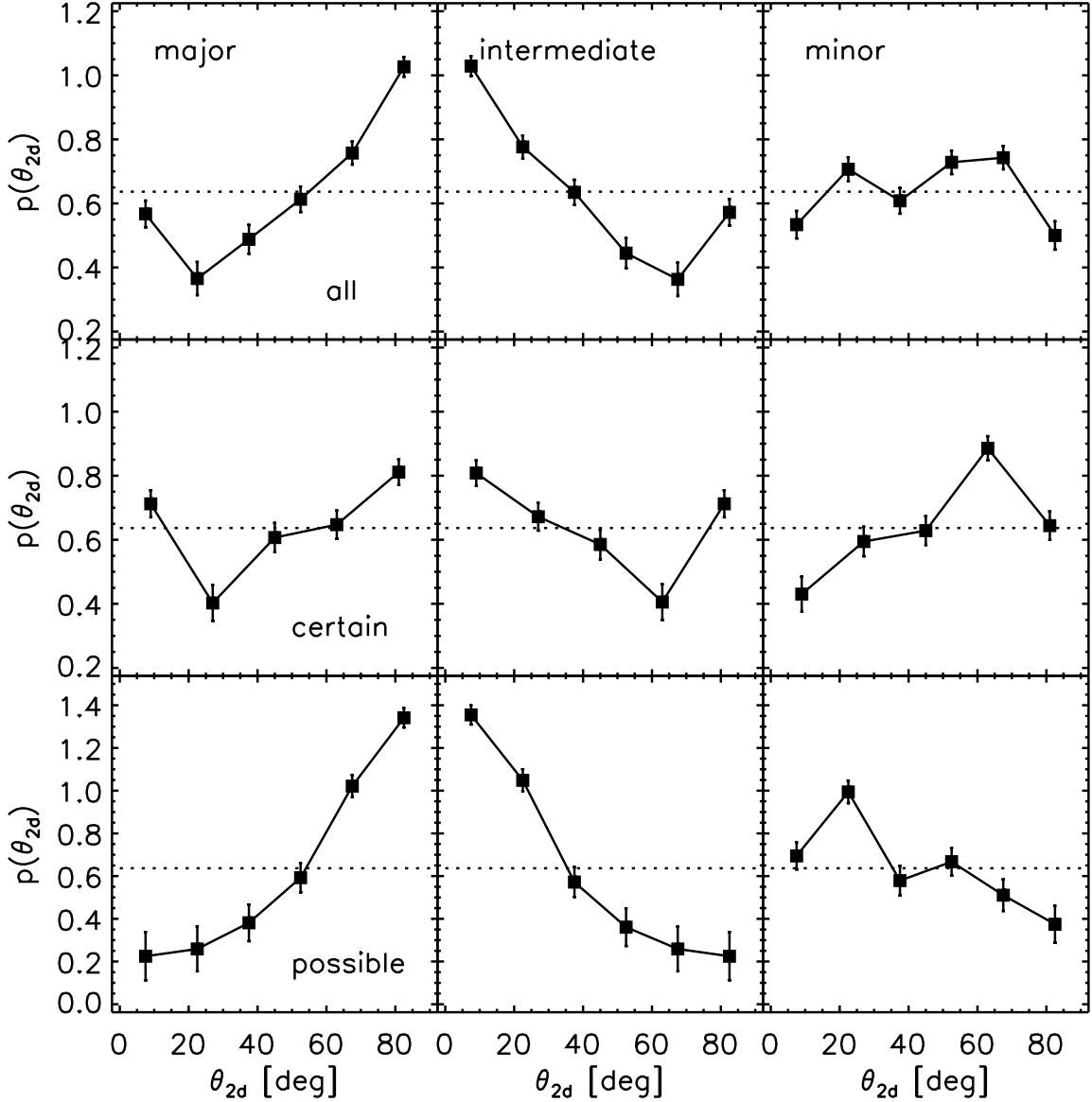}
\caption{Probability density distributions of the angles between the position vectors of the Virgo  galaxies and the major, 
intermediate, and minor principal axes of the local velocity shear tensors in the left, middle, and right panel, respectively. 
The principal axes and the position vectors are all projected onto the two dimensional sky plane orthogonal to the line-of-sight 
direction to the Virgo center. The positions of the Virgo galaxies are measured relative to the Virgo center, M87. 
The top, middle, and bottom panel corresponds to the case of all, certain, and possible members of the Virgo cluster, 
respectively. }
\label{fig:all}
\end{center}
\end{figure}

Figure \ref{fig:all} shows  $p(\theta_{2d})$ with the Poisson errors versus $\theta_{2d}$ in unit of degree for the case of all, certain, and 
possible members of the Virgo cluster in the top, middle, and bottom panel, respectively. The left, middle, and right panel corresponds 
to the alignment angles with the major, intermediate, and minor principal axis of the local velocity shear, respectively. 
In each panel, the dotted line represents the uniform distribution of $p(\theta_{2d})=2/\pi$, corresponding to the case of no 
alignment. The error,  $\sigma$, of each $\theta_{2d}$-bin is calculated as one standard deviation of the Poisson variable as 
$\sigma = 2/(\pi\sqrt{N_{\theta}-1})$ where $N_{\theta}$ represents the number of the EVCC galaxies belonging to each 
$\theta_{2d}$-bin. As can be seen,  the projected two dimensional positions of the EVCC galaxies  have a strong tendency of being 
aligned with (perpendicular to) the projected intermediate (major) principal axis of the local velocity tensor.

Since both of the real three dimensional major and intermediate principal axes of the local velocity tensor almost lie in the plane 
of the sky, the projected major and intermediate principal axes in the plane of the sky are orthogonal to each other. Therefore, 
the probability density distributions $p(\theta_{2d})$ of the middle panels are in fact the same as $p(\pi/2-\theta_{2d})$ of 
the left panels. The right panels of Figure \ref{fig:all} show that there is no signal of alignment between the EVCC galaxies 
and the projected minor principal axis, which just reflects the fact that the real direction of the minor principal axis is completely 
lost when projected. Henceforth, what Figure \ref{fig:all} essentially tells us is that the positions of the EVCC galaxies tend to 
be aligned with the intermediate principal axis of the local tidal shear.

Figure \ref{fig:all} also reveals that the possible members of the Virgo cluster are much more strongly aligned with the intermediate  
principal axis of the local velocity shear than the certain members. Obviously the relaxation process inside the Virgo cluster 
must have an effect of erasing the memory of the member galaxies about the anisotropic merging along the principal axes of the 
local velocity shear.  Expecting that the possible members in the EVCC are located in the cosmic filament that interconnects the 
Virgo cluster with the surrounding matter distribution, the strong signal of the alignment detected in the bottom panel of 
Figure \ref{fig:all} provides an observational evidence for the recent claim of \citet{libeskind-etal13a} and \citet{tempel-etal14} 
that the velocity shear field is a good tracer of the cosmic web. 

\begin{figure}[ht]
\begin{center}
\epsscale{1.0}
\plotone{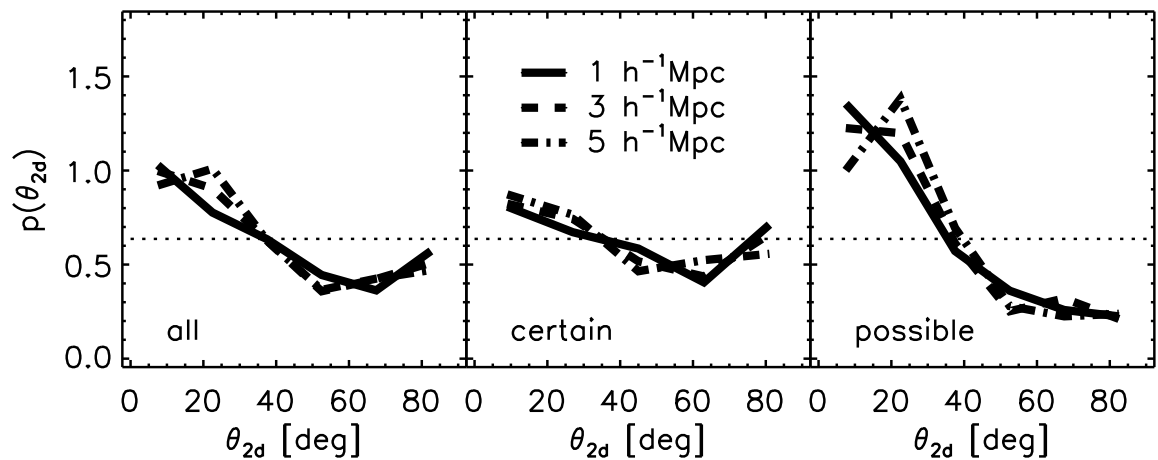}
\caption{Probability density distributions of the angles of the alignments between the projected positions of the EVCC 
galaxies and the projected intermediate principal axes of the local velocity shear tensor for three different cases of the 
filtering radius, $R_{f}$: In each panel, the solid, dashed, and dot-dashed line corresponds to $R_{f}=1,\ 3,$ and $5$ in 
unit of $h^{-1}$Mpc, respectively.}
\label{fig:filter}
\end{center}
\end{figure}

We investigate how the strength of the alignment signal changes with the smoothing scale, by repeating the 
determination of $p(\theta_{2d})$ for three different cases of $R_{f}$. 
Figure \ref{fig:filter} show the probability density distribution of the alignment angles between the intermediate principal axes 
of the velocity shear tensor and the position vectors of all, certain, and possible members of the EVCC in the 
left, middle, and right panel, respectively. In each panel, the solid, dashed, and dot-dashed line corresponds to the case of 
$R_{f}=1,\ 3$ and $5\,h^{-1}$Mpc, respectively. As can be seen, no evolution of the alignment signal with the smoothing scale has been 
found, which is consistent with the recent numerical finding of \citet{libeskind-etal14} that the directions of the principal axes of the the 
velocity shear field are quite robust against the change of the smoothing scale.

\begin{figure}[ht]
\begin{center}
\epsscale{1.0}
\plotone{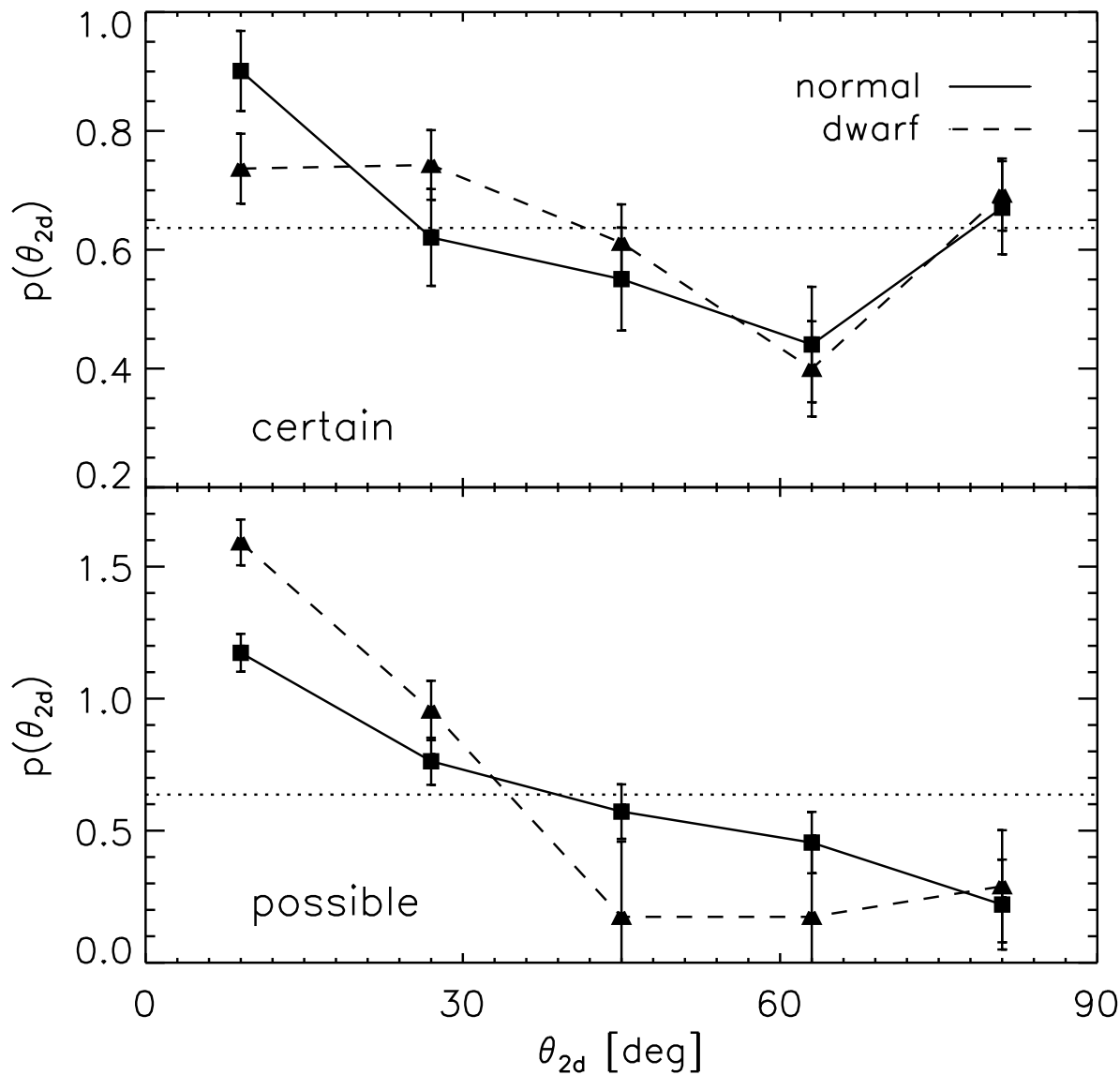}
\caption{Probability density distributions of the angles of the alignments between the projected positions of the certain 
(possible) members and the intermediate principal axis of the local velocity shear in the top (bottom) panel. 
In each panel, the solid and the dashed line corresponds to the case of the dwarf and the normal galaxies, respectively. }
\label{fig:mass}
\end{center}
\end{figure}

To see if and how the strength of the anti-alignment signal varies with the masses of the EVCC galaxies, we determine 
$p(\theta_{2d})$ for the low-mass dwarf and for the high-mass normal galaxies separately. A total of $510$ galaxies in the subsample 
of the certain members are found to be classified as dwarfs while the subsample of the possible members contains a total of $110$ 
dwarf galaxies \citep[see][for the details of the morphology classification]{kim-etal14}. 
In each panel of Figure \ref{fig:mass}, the solid and the dashed line corresponds to $p(\theta_{2d})$ for the dwarf and for the 
normal galaxies, respectively.   The smoothing scale for the local velocity tensor is reset again at $R_{f}=1\,h^{-1}$Mpc. 
As can be seen, for the case of the certain members (top panel) there is almost no difference in the strength of the alignment 
signal between the dwarf and the normal galaxies. Whereas, for the case of the possible members (bottom panel) the dwarf galaxies 
yield much stronger signal of alignment than the normal galaxies. 

Our physical explanation for this {\it apparent} mass-dependence of the alignment strength is as follows.  For the case of the normal galaxies, 
the events of their merging into the Virgo cluster occur preferentially along the local filament. Provided that the local filament is aligned with the 
minor principal axis of the local velocity shear parallel to the line of sight direction, the alignment tendency of the normal satellites 
with the minor principal axis of the local velocity shear tensor would be significantly lost by the projection onto the plane of sky.  
Whereas, for the case of the dwarf galaxies, the merging events occur in the local sheet that encompasses the Virgo cluster and 
the local filament. Provided that the local sheet is parallel to the plane spanned by the intermediate and minor principal axis of the local 
velocity shear, the projected positions of the dwarf satellites would still appear to be strongly aligned with the intermediate principal axis 
of the local velocity shear tensor. 

\section{SUMMARY AND CONCLUSION}\label{sec:con}

In the field of the large scale structure of the universe, one of the most urgent issues to address is how to trace and quantify the 
cosmic web whose ubiquity and coherence has been noted for long ever since \citet{web96} attempted to explain its presence in the 
context of the CDM paradigm.  In the light of the works of \citet{hoffman-etal12}, \citet{libeskind-etal12}, and \citet{libeskind-etal13a} 
which proposed a web-classification scheme based on the velocity shear tensor  and demonstrated by $N$-body simulations that the 
principal axes of the velocity shear tensors are well aligned with the directions of the cosmic web, we have performed an observational 
test of their numerical prediction by measuring the anisotropic spatial distributions of the galaxies in and around the Virgo cluster in 
the principal frame of the local velocity shear tensor. 

The Virgo cluster has been chosen as an optimal target for this observational test because of its proximity and young dynamical 
state.  The positions of the galaxies in and around the Virgo cluster have been extracted from the EVCC \citep{kim-etal14} and the local 
velocity tensor at the Virgo center has been estimated by analyzing the peculiar velocity field reconstructed from the SDSS DR7 
\citep{wang-etal12}.  To avoid the contamination caused by the nonlinear redshift distortion effect on the positions of the EVCC 
galaxies, the measurements have been made in the projected two dimensional plane of the sky orthogonal to the line of sight direction 
to the Virgo center. The direction of the minor principal axis of the local velocity shear has turned out to be completely lost by the projection 
since it has been found to be almost perpendicular to the plane of the sky, while the major and the intermediate principal axes have been 
found to be only slightly modified by the projection onto the plane of the sky. 

A clear signal of alignment has been detected between the projected positions of the EVCC galaxies and the projected 
intermediate principal axis of the local velocity shear.  The signal has been found to be robust against the change of the smoothing scale 
from $1$ to $5\,h^{-1}$Mpc, which is consistent with the recent numerical result of \citet{libeskind-etal14} that the directions of the 
principal axes of the velocity shear field are insensitive to the smoothing scales. Those galaxies classified by Kim et al. (2014) as possible 
members according to their radial velocities have been found to be more strongly aligned than those galaxies classified as certain members. 
This result indicates that after merging into a host cluster the positions of the satellites become randomized relative to the principal 
frame of the local velocity shear, providing an observational support for the previous numerical result based on $N$-body simulations 
that the subhalos fallen earlier tend to be less strongly aligned with the elongated directions of the shapes of the host halos than 
those just fallen \citep[e.g.,][and references therein]{wang-etal14}. 

Interestingly, the dwarf satellites around the Virgo cluster have been shown to be more strongly aligned with the principal axis of the 
local velocity shear tensor than the normal counterparts. Our interpretation is that this difference in the alignment strength between the normal 
and the dwarf satellites stems from the difference in the infall direction between them. The normal galaxies merge into the Virgo cluster preferentially 
along the local filament, while the infall of dwarf galaxies occurs in the local sheet that encompasses the Virgo cluster and the local filament.  
Supposing that the local filament is aligned with the minor principal axis of the local velocity shear tensor,  the normal satellites are expected to be 
aligned with the minor principal axis parallel to the line of sight direction, so that the projection would diminish their alignment tendency. 
Whereas, the projection would not affect much the alignment tendency of the dwarf satellites with the intermediate axis of the local 
velocity shear tensor provided that the local sheet is parallel to the plane spanned by the minor and intermediate principal axis perpendicular to the 
line of sight direction. Henceforth, our conclusion is that the local sheet where the dwarf satellites of the Virgo cluster reside is well aligned with the 
principal directions of the local velocity shear , supporting the recent numerical claim of \citet{libeskind-etal13a} and \citet{tempel-etal14} that the velocity 
shear field is a good tracer of the cosmic web. 

In the current work, we have used the {\it linearly reconstructed} velocity shear field  to investigate the alignments between the 
Virgo satellites and their principal directions. 
Since the linear reconstruction assumes that the velocity shear field is curl free, our analysis has completely neglected the presence of the 
vorticity. However, a couple of recent literatures proved that the vorticity effect on the cosmic web alignments 
becomes important in the nonlinear regime  \citep{libeskind-etal13b,libeskind-etal14}. We speculate that the central member galaxies of the
Virgo cluster which have shown only weak signal of the alignment with the intermediate principal axis of the velocity shear field in the 
current work might be aligned with the direction of the vorticity.  To observationally test this speculation, the nonlinear reconstruction of 
the peculiar velocity field will be required, which is beyond the scope of this paper.

It will be also interesting to explore how the galaxies in and around the Coma cluster are aligned with the principal axes of its 
local velocity shear tensor and to compare the result with our current result. 
The Coma cluster located in the filamentary Coma Supercluster \citep{GT78} has been known to be in a dynamically unrelaxed state 
like the Virgo cluster \citep[e.g.,][]{briel-etal92}. However, an observation has shown that not only the galaxies in the Coma cluster but 
also the galaxies in the Coma Supercluster show no preferential directions in their spatial orientation \citep{torlina-etal07}. 
It might be due to a degeneracy among the three principal axes of the velocity shear tensor at the Coma site. We plan to work on these 
interesting topics and hope to report the results in the future.

\acknowledgments

We thank an anonymous referee for helping us significantly improve the original manuscript.  We also thank H. Wang for the 
peculiar velocity data and Aeree Chung for useful comments.  We are very grateful to H. Jerjen, T. Lisker, E.-C. Sung, Y. Lee, J. 
Chung, M. Pak, W. Yi, and W. Lee for their valuable contribution to the construction of the EVCC. 
This work was supported by the research grant from the National Research Foundation of Korea to the Center for 
Galaxy Evolution Research  (NO. 2010-0027910).  JL also acknowledges the financial support by the Basic Science Research 
Program through the National Research Foundation of Korea (NRF) funded by the Ministry of Education (NO. 2013004372).
SCR was supported by Basic Science Research Program through the NRF funded by the Ministry of Education 
(NRF-2012R1A1B4003097).  SK acknowledges support from the National Junior Research Fellowship of NRF (No. 2011-0012618).

\clearpage
\clearpage
\clearpage
\clearpage
\clearpage
\end{document}